\documentclass{aa}
\usepackage{natbib}
\usepackage{graphicx}
\usepackage{txfonts}
\usepackage{xcolor}
\usepackage{float}
\usepackage{comment}
\usepackage{pgf}   
\usepackage{siunitx}
\usepackage{transparent}
\DeclareSIUnit \jansky {Jy}
\usepackage{makecell}

\newcommand\gsim{\mathrel{\hbox{\rlap{\lower.55ex \hbox {$\sim$}}
                   \kern-.3em \raise.4ex \hbox{$>$}}}}
\newcommand\lsim{\mathrel{\hbox{\rlap{\lower.55ex \hbox {$\sim$}}
                   \kern-.3em \raise.4ex \hbox{$<$}}}}

\begin{document} 

\title{Satellite shadows through stellar occultations} \author{Paul J. Groot \inst{1,2,3}} 
   \institute{
   Department of Astrophysics/IMAPP, Radboud University, P.O.Box 9010, 6500 GL Nijmegen, The Netherlands\\
   \email{p.groot@astro.ru.nl}
   \and 
    South African Astronomical Observatory, PO Box 9, Observatory, 7935, Cape Town, South Africa \and
    Department of Astronomy, University of Cape Town, Private Bag X3, Rondebosch, 7701, South Africa}
             
\date{Received month day, year; accepted month day, year}


   \abstract
  {}
{    The impact of mega-constellations of satellites in low-Earth orbit
    during nighttime optical observations is assessed.}
   {Orbital geometry is used to calculate the impact of stellar
     occultations by satellites on the photometry of individual stars
     as well as the effect on the photometric calibration of
     wide-field observations.}
   {Starlink-type satellites will have occultation disks several
     arcseconds across. Together with occultation crossing times of
     0.1--100 msec, this will lead to photometric `jitter' on the flux
     determination of stars. The level of impact for a given star
     depends on the ratio of the integration time of the frame over
     the occultation crossing time. In current-day CCD-based synoptic
     surveys, this impact is negligible ($<< 1\%$), but with
     future CMOS-based wide-field surveys obtaining data at
     frequencies $>$1Hz, the impact will grow towards complete
     dropouts. At integration times similar to the occultation
     crossing time, the orbit of a satellite can be traced using the
     occultation method. At even shorter integration times, the shape
     of the occulting satellite can be deduced. }
  {Stellar occultations by passing satellites, enabled by high-speed
     CMOS technology, will be a new method for studying orbiting
     satellites. Large-scale monitoring programs will be needed to
     independently determine and update the orbits of
     satellites. }


\titlerunning{Satellite shadows through stellar occultations}
\authorrunning{Groot, P.J.}
\maketitle
%
\section{Introduction}
As mega-constellations of man-made satellites are being launched,
their impact on astronomy needs to be assessed. The aspect of
pollution in the optical and radio bands has been greatly emphasised (e.g. \citealt{Witze19, Galozzi20}). In the optical, this pollution is
mostly restricted to the twilight hours of the night when the
satellites still catch sunlight while night has already descended on
the ground-based telescopes (see e.g. \citealt{Bassa22, Mroz22,
  TregloanReed21, Mallama21, Tyson20, Hainaut20, McDowell20}). However,
the effect of satellite constellations on optical ground-based
observations is not limited to the twilight hours. During the night,
satellites will cause occultations of stars located along their
path. As both the occulter (the satellite) and the occulted object
(the star) are above the Earth's atmosphere, the occultation will be
complete as the angular diameter of the satellite ($\theta_s$) will be
much larger than that of the background star ($\theta_\star$).

In Sect.\ \ref{sec:basics} we revisit the basic geometries
involved. In Sect.\ \ref{sec:constellations} we overview various
current and upcoming satellite constellations.  In
Sect.\ \ref{sec:impact} we apply the analysis to a number of satellite
types, and in Sect.\ \ref{sec:outlook} we revisit the question on how
feasible it is to discover and characterise satellites through optical
observations of stellar occultations.
 
 \section{Geometry \label{sec:basics}} 
 Throughout this exercise we make two simplifying
 assumptions. First, we assume a satellite has a circular shape
 as seen from any observatory on Earth, characterised by a satellite
 radius, $R_s$, over which area the satellite is completely
 opaque. Secondly, we assume the path of the satellite to be on a
 great circle passing through the zenith.
 
 The assumed geometry is that of a satellite circling the Earth (with
 radius, $R_\oplus$, taken as 6378 km) at an orbital altitude $h$. With
 an Earth mass, $M_\oplus$, taken as 5.976$\times$10$^{24}$ kg, the
 orbital period, $P$, follows from Kepler's laws as
 \begin{equation}
     P = \sqrt{\frac{4\pi^2a_s^3}{GM_{\oplus}}} = 2\pi \sqrt{\frac{a_s^3}{GM_{\oplus}}}, 
 \end{equation}
 with $a_s$ = $R_\oplus+h$, the radius of the satellite's orbit from
 the centre of the Earth, and $G$ Newton's gravitational constant.
 
 The tangential velocity of a satellite passing overhead,
   ignoring the Earth's rotation, is given by
 \begin{equation}
    v_s = \frac{2\pi a_s}{P} = \sqrt{\frac{G M_\oplus}{a_s}}. 
\end{equation}
 The satellite will produce an occultation disk ($\theta_s$) with an angular
 diameter of
 \begin{equation}
 \theta_s = {\rm atan}{(\frac{2R_s}{h})} \simeq \frac{2R_s}{h},
 \end{equation}
 where the approximation is justified for the small angles involved. The angular diameter of a satellite is, in all practical cases, much
larger than the angular diameter ($\theta_{\star}$) of a background star ($\theta_{\star}
<< 10\,$ mas). The occultation will therefore be complete, and the
duration is set by the rate of motion of the satellite. To first
order, the crossing time, $\tau_s$, is the time it takes the satellite
to move its own size as it passes through the zenith, that is,
\begin{equation}
    \tau_s = \frac{2 R_s}{v_s},
\end{equation}
where $v_s$ is the orbital velocity of the satellite. It is the crossing time, $\tau_s$, that needs to be compared to the
typical integration time in synoptic surveys to assess the impact of a
stellar occultation by a satellite.

The size of the satellite's shadow on Earth is equal to the size of
the satellite, as stars can be taken to be infinitely far away. Hence,
if the aperture, $D$, of the telescope is larger than the size of the
satellite ($D>2R_s$), the occultation will not be total and the
ingress and egress duration of the occultation will be flattened. This
sets an upper limit to the size of the telescope used if one aims to detect a total occultation.

\section{Shadows by satellite constellations \label{sec:constellations}}

Most satellites can be grouped into a small number of orbits:
low-Earth orbit (LEO), at altitudes $h$=350-500 km; the global
positioning system (GPS) medium-Earth orbit (MEO) belt at
$h\sim$20\,000 km; and geostationary (GEO) satellites at $h$=36\,000
km. The sizes of satellites vary strongly, from CubeSats with $R_s
\sim 0.5$m, to Starlink-type satellites with $R_s \sim 5$m, to larger,
but rarer, satellites such as the KeyHole Hubble-type satellites with
$R_s \sim 15$m and the Orion/Mentor radio dishes ($R_s \sim 100$m) in
GEO orbits.

\begin{figure}[htb]
  \centering
  \includegraphics[width=1.0\linewidth]{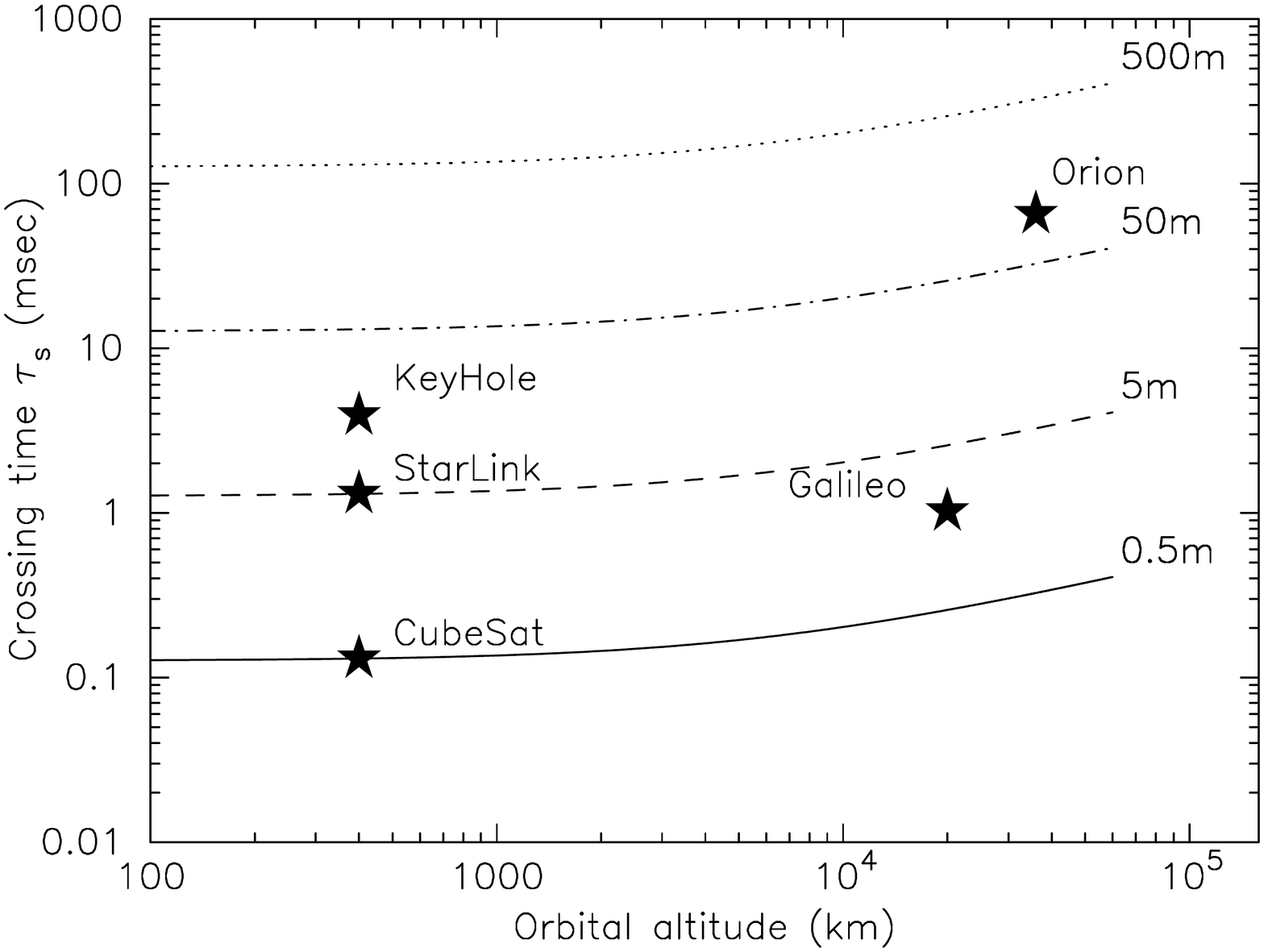}
 \caption[]{Shadow crossing time as a function of orbital altitude of a satellite around Earth, for various satellite sizes. Indicated are representative classes of satellites in LEO, MEO, and GEO orbits.
    \label{fig:crossing}}
\end{figure}

In Table\ \ref{tab:crossing} we list the relevant occultation disks
and crossing times for the categories listed
above. Figure\ \ref{fig:crossing} shows the crossing times as a
function of orbital altitude and size of the satellite.
Figure \ref{fig:crossing} shows that, for feasible satellite sizes, the
crossing times are generally $\lsim$100 msec. Although this appears
low, we highlight the impact of this in Sect.\ \ref{sec:impact}.

\begin{table}[htb]
\caption[]{Overview of orbital altitudes ($h$), sizes ($R_s$), angular diameters ($\theta_s$), speeds ($\omega_s$), and crossing times ($\tau_s$) of satellite categories \label{tab:crossing}}
\begin{tabular}{lrrrrr}
  \hline \hline
Orbit/Class & $h$ & $R_s$ & $\theta_s$ & $v_s$ & $\tau_s$\\
& [km] & [m] & [\arcsec] & [km/s] & [msec] \\ \hline
LEO `CubeSat'   & 400       & 0.5   & 0.5   & 7.67 & 0.13\\
LEO `Starlink'  & 400       & 5     & 5.2   & 7.67 & 1.30\\
LEO `KeyHole'   & 400       & 15    & 15.5  & 7.67 & 3.91\\
GPS `Galileo'   & 20\,000   & 2     & 0.04  & 3.89  & 1.02\\
GEO `Orion'     & 36\,000   & 100   & 1.14  & 3.06  & 65.2\\
\hline
\end{tabular}
\end{table}

\section{Impact on ground-based astronomy \label{sec:impact}}

Mega-constellations of satellites are of major concern to ground-based
astronomy. With more than 40\,000 satellites planned to be launched
over the next few years, ground-based astronomy will face the
situation that there will be one to several satellites at any moment
per square degree of sky (see e.g. \citealt{Bassa22}). Major synoptic
surveys with instantaneous fields-of-view, $\Omega$ $>$ 1 square
degree, will have multiple to tens of satellites crossing each image
\citep{Mroz22}. The effect on nighttime ground-based photometry will
be that occulted stars will be completely blocked out for a crossing
time, $\tau_s$. The effect on the resulting photometry depends on the
ratio of the crossing time over the integration time, $t_{\rm int}$,
of an exposure: $f = t_{\rm int} / \tau_s$. The impact depends on
whether this ratio, $f$, is larger or smaller than unity.

\subsection{Long integration times: $f>1$ \label{sec:longint}}
Most current ground-based synoptic surveys use integration times
$t_{\rm int} > 1$s. The effect of a satellite shadow passing through
the field-of-view is therefore that of an occulting trail that blocks
part of the integrated light of stars located along its path with a
width $\theta_s$. The fractional decrease in the light of any given
star is equal to $1/f$, assuming the occultation to be complete for
the crossing time, $\tau_s$, which also depends on the ratio $2R_s/D$.

For a Starlink-type satellite, the ratio $f$ is $\sim$12\,000 for a
Vera Rubin Observatory exposure time of 15s \citep{lsst17}, $\sim$23\,000 for
Zwicky Transient Facility integration times of 30s \citep{ztf19},
and $\sim$46\,000 for BlackGEM exposure times of 60s
\citep{Bloemen2016}. As the absolute photometric precision in
ground-based astronomy rarely exceeds the 1\% level, such drops in the
light curve may not be noticeable as of yet.

The area affected, $f_{\Omega}$, by a worst-case crossing (corner to corner) is about
$f_{\Omega}\sim$ 0.2\% (= $\sqrt{2}\times \theta_s/(3600*A)$) for a
wide-field imager with an $A \times A$ square degrees field-of-view. In
some parts of the Galactic plane and bulge, the ground-based,
  discernible stellar densities reach $>$ 500\,000 per square degree
down to $g\sim$ 20 mag \citep{Drew2014_VPHAS, Groot2009_UVEX},
and therefore hundreds of stars will lie in the path of occultation.

A secondary effect is introduced by the absolute photometric
calibration of a single frame. Generally, this is achieved by measuring
the flux of a number of calibrator stars across the field-of-view and
taking the ratio of the measured flux with the tabulated brightness of
these stars. This is currently done using \textit{Gaia} Data Release 2 or 3 (DR2 or DR3)
photometry for the calibrator stars. As a satellite shadow crosses the
field-of-view during an integration, the measured brightness of the
calibrator stars that lie on the path of the satellite will be lower
than expected, and therefore the photometric calibration of the whole
frame will be affected, artificially `brightening' the whole frame: as
the calibrators are fainter than expected, this will result in an
artificial increase in the apparent brightness of all other stars in
the frame. In current-day surveys this will not be a major problem as
the calibrator dimming will be equal to the ratio of the number of
affected (calibrator) stars over the total number of (calibrator)
stars in the image. For a 1\degr $\times$ 1\degr\, field-of-view and a
single Starlink-type satellite, the occulted area is, at worst, 0.2\%
of the total area (7.2 square arcminutes), and therefore,
assuming a uniform stellar density, also only 0.2\% of stars are
affected, at a level of 1/$f$ per star, so the total impact on the
photometric calibration is negligible.

We note that the total crossing time, $t_{\rm cross}$, of a Starlink-type satellite over
a 1-degree field-of-view is equal to $t_{\rm cross}$ =
${A}/{\theta_s\tau_s}$ = 0.9s. To demonstrate these satellite shadows,
the best current-day option would be to observe GEO
satellites that will cause a trail of 1 arcminute for a 60s exposure.

\subsection{Short integration times: $f \lsim 1$ \label{sec:shortint}}
CMOS-camera technology is rapidly advancing. CMOS cameras with frame
sizes exceeding 6\,000 x 6\,000 pixels and quantum efficiencies
$>$90\% are available. The expectation is therefore that within a few
years CMOS technology will completely replace current-day CCD
technology in ground-based astronomy (e.g. \citealt{KISO22,Zhang20}). With this comes the possibility to take
much shorter exposures thanks to the parallel read-out technology in CMOS
cameras. Integration times shorter than 1s become feasible. As can be
seen in Table\ \ref{tab:crossing} and Fig.\ \ref{fig:crossing}, the 1\%
influence on any given star in absolute photometry becomes relevant
for integration times shorter than 0.1s for Starlink-type satellites,
and shorter than 6s for an Orion-type satellite.

Full-frame read-out speeds on large-format CMOS cameras are currently
in the 24Hz range. Therefore, `SuperHertz' wide-field cameras will
appear, coupled to large-aperture telescopes (e.g. \citealt{KISO18,KISO22}).

At 24Hz the integration times come in the range of the crossing times given in
Table\ \ref{tab:crossing}. With 2$\times$2 binning, speeds up to 100Hz
can be achieved, and a 10\% dimming effect can be expected for
Starlink-type satellites. With further binning or windowing crossing
times can be resolved in time, and the path of a satellite can be
followed by a series of `dropouts' of stars along the orbit. This
will allow for orbit determinations of satellites larger than a few
metres, depending on their orbital altitudes.

Additionally, the assumption of a circular, fully opaque occultation
patch can be relaxed. Set against a sufficiently dense stellar field,
the shape of each satellite can be deduced from the complex light
curve that will be produced as the satellite path is followed across
the sky. With Gaia (Early) DR3 astrometric precision ($\sigma$) on the
location of each star at a level of 1 mas \citep{GaiaEDR3Ast21}, the
ratio $\sigma/\theta_s << f$, and therefore it will be the integration time,
$t_{\rm int}$, that is the limiting factor on the level of detail that
will be retrievable. Conceptually, the level of detail will be of the
order of $f R_s$.

For satellites in {known} orbits CMOS cameras can be sub-arrayed
to allow for read-out speeds of $\sim$1000 Hz, leading to smaller $f$. A
detailed modelling is outside the scope of this study, but can easily
be conceptualised. This method will allow for the determination of the
position and shape of a satellite at accuracies \mbox{$<$1 m} when $f$
reaches 0.1 or less.

The effect noted in Sect.\ \ref{sec:longint} on the photometric
calibration of the full frame will disappear when $f<1$. Although the
effect on each individual star will become stronger as $f$ approaches
unity, the length of a trail through each exposure will become shorter
and shorter, and $f_\Omega$ will approach zero.

\section{Outlook and discussion \label{sec:outlook}}

Mega-constellations in LEO will have a significant impact
on ground-based astronomy, in particular for wide-field synoptic
surveys. During twilight hours the reflected sunlight will cause
bright streaks in the images. We have shown that during hours of
darkness, satellites will cause a shadow path by blocking out the
light of background stars through occultations.

For current-day longer-duration exposures ($t_{\rm int} >$ 1 s), the
effect is modest and restricted to downwards photometric jitter on
individual stars. However, with CMOS technology advancing quickly,
integration times similar to the satellite's crossing time over any
given star come within reach. This will lead to strong dropouts on
an individual star's light curve. It also opens up the possibility to
use the occultation trail to detect and track the satellite in its
orbit across the sky. This effect is completely independent from the
albedo or `stealth' capabilities of the satellite, and also extends
beyond just the optical wavelength regime.

With integration times shorter than the occultation crossing time, and
set against sufficiently crowded stellar fields (e.g. the Galactic
plane and bulge, or the Magellanic Clouds), the shape of satellites will be
deducible, opening up a completely new way to characterise satellites
in orbit.

\begin{acknowledgements}
PJG is partially supported by NRF SARChI grant 111692. The author
thanks Frank Verbunt for an insightful discussion, and David Palmer
for pointing out a conceptual mistake in a previous version and making
useful suggestions for simplifications.
\end{acknowledgements}


\begin{thebibliography}{18}
\expandafter\ifx\csname natexlab\endcsname\relax\def\natexlab#1{#1}\fi

\bibitem[{{Bassa} {et~al.}(2022){Bassa}, {Hainaut}, \&
  {Galad{\'\i}-Enr{\'\i}quez}}]{Bassa22}
{Bassa}, C.~G., {Hainaut}, O.~R., \& {Galad{\'\i}-Enr{\'\i}quez}, D. 2022,
  \aap, 657, A75

\bibitem[{{Bellm} {et~al.}(2019){Bellm}, {Kulkarni}, {Barlow}, {Feindt},
  {Graham}, {Goobar}, {Kupfer}, {Ngeow}, {Nugent}, {Ofek}, {Prince}, {Riddle},
  {Walters}, \& {Ye}}]{ztf19}
{Bellm}, E.~C., {Kulkarni}, S.~R., {Barlow}, T., {et~al.} 2019, \pasp, 131,
  068003

\bibitem[{{Bloemen} {et~al.}(2016){Bloemen}, {Groot}, {Woudt}, {Klein Wolt},
  {McBride}, {Nelemans}, {K{\"o}rding}, {Pretorius}, {Roelfsema}, {Bettonvil},
  {Balster}, {Bakker}, {Dolron}, {van Elteren}, {Elswijk}, {Engels}, {Fender},
  {Fokker}, {de Haan}, {Hagoort}, {de Hoog}, {ter Horst}, {van der Kevie},
  {Koz{\l}owski}, {Kragt}, {Lech}, {Le Poole}, {Lesman}, {Morren}, {Navarro},
  {Paalberends}, {Paterson}, {Paw{\l}aszek}, {Pessemier}, {Raskin}, {Rutten},
  {Scheers}, {Schuil}, \& {Sybilski}}]{Bloemen2016}
{Bloemen}, S., {Groot}, P., {Woudt}, P., {et~al.} 2016, in Society of
  Photo-Optical Instrumentation Engineers (SPIE) Conference Series, Vol. 9906,
  Ground-based and Airborne Telescopes VI, ed. H.~J. {Hall}, R.~{Gilmozzi}, \&
  H.~K. {Marshall}, 990664

\bibitem[{{Drew} {et~al.}(2014){Drew}, {Gonzalez-Solares}, {Greimel}, {Irwin},
  {K{\"u}pc{\"u} Yoldas}, {Lewis}, {Barentsen}, {Eisl{\"o}ffel}, {Farnhill},
  {Martin}, {Walsh}, {Walton}, {Mohr-Smith}, {Raddi}, {Sale}, {Wright},
  {Groot}, {Barlow}, {Corradi}, {Drake}, {Fabregat}, {Frew}, {G{\"a}nsicke},
  {Knigge}, {Mampaso}, {Morris}, {Naylor}, {Parker}, {Phillipps}, {Ruhland},
  {Steeghs}, {Unruh}, {Vink}, {Wesson}, \& {Zijlstra}}]{Drew2014_VPHAS}
{Drew}, J.~E., {Gonzalez-Solares}, E., {Greimel}, R., {et~al.} 2014, \mnras,
  440, 2036

\bibitem[{{Gallozzi} {et~al.}(2020){Gallozzi}, {Paris}, {Scardia}, \&
  {Dubois}}]{Galozzi20}
{Gallozzi}, S., {Paris}, D., {Scardia}, M., \& {Dubois}, D. 2020, arXiv
  e-prints, arXiv:2003.05472

\bibitem[{{Groot} {et~al.}(2009){Groot}, {Verbeek}, {Greimel}, {Irwin},
  {Gonz{\'a}lez-Solares}, {G{\"a}nsicke}, {de Groot}, {Drew}, {Augusteijn},
  {Aungwerojwit}, {Barlow}, {Barros}, {van den Besselaar}, {Casares},
  {Corradi}, {Corral-Santana}, {Deacon}, {van Ham}, {Hu}, {Heber}, {Jonker},
  {King}, {Knigge}, {Mampaso}, {Marsh}, {Morales-Rueda}, {Napiwotzki},
  {Naylor}, {Nelemans}, {Oosting}, {Pyrzas}, {Pretorius}, {Rodr{\'\i}guez-Gil},
  {Roelofs}, {Sale}, {Schellart}, {Steeghs}, {Szyszka}, {Unruh}, {Walton},
  {Weston}, {Witham}, {Woudt}, \& {Zijlstra}}]{Groot2009_UVEX}
{Groot}, P.~J., {Verbeek}, K., {Greimel}, R., {et~al.} 2009, \mnras, 399, 323

\bibitem[{{Hainaut} \& {Williams}(2020)}]{Hainaut20}
{Hainaut}, O.~R. \& {Williams}, A.~P. 2020, \aap, 636, A121

\bibitem[{{Lindegren} {et~al.}(2021){Lindegren}, {Klioner}, {Hern{\'a}ndez},
  {Bombrun}, {Ramos-Lerate}, {Steidelm{\"u}ller}, {Bastian}, {Biermann}, {de
  Torres}, {Gerlach}, {Geyer}, {Hilger}, {Hobbs}, {Lammers}, {McMillan},
  {Stephenson}, {Casta{\~n}eda}, {Davidson}, {Fabricius}, {Gracia-Abril},
  {Portell}, {Rowell}, {Teyssier}, {Torra}, {Bartolom{\'e}}, {Clotet},
  {Garralda}, {Gonz{\'a}lez-Vidal}, {Torra}, {Abbas}, {Altmann}, {Anglada
  Varela}, {Balaguer-N{\'u}{\~n}ez}, {Balog}, {Barache}, {Becciani}, {Bernet},
  {Bertone}, {Bianchi}, {Bouquillon}, {Brown}, {Bucciarelli}, {Busonero},
  {Butkevich}, {Buzzi}, {Cancelliere}, {Carlucci}, {Charlot}, {Cioni},
  {Crosta}, {Crowley}, {del Peloso}, {del Pozo}, {Drimmel}, {Esquej}, {Fienga},
  {Fraile}, {Gai}, {Garcia-Reinaldos}, {Guerra}, {Hambly}, {Hauser},
  {Jan{\ss}en}, {Jordan}, {Kostrzewa-Rutkowska}, {Lattanzi}, {Liao}, {Licata},
  {Lister}, {L{\"o}ffler}, {Marchant}, {Masip}, {Mignard}, {Mints}, {Molina},
  {Mora}, {Morbidelli}, {Murphy}, {Pagani}, {Panuzzo}, {Pe{\~n}alosa Esteller},
  {Poggio}, {Re Fiorentin}, {Riva}, {Sagrist{\`a} Sell{\'e}s}, {Sanchez
  Gimenez}, {Sarasso}, {Sciacca}, {Siddiqui}, {Smart}, {Souami}, {Spagna},
  {Steele}, {Taris}, {Utrilla}, {van Reeven}, \& {Vecchiato}}]{GaiaEDR3Ast21}
{Lindegren}, L., {Klioner}, S.~A., {Hern{\'a}ndez}, J., {et~al.} 2021, \aap,
  649, A2

\bibitem[{{LSST Science Collaboration} {et~al.}(2017){LSST Science
  Collaboration}, {Marshall}, {Anguita}, {Bianco}, {Bellm}, {Brandt},
  {Clarkson}, {Connolly}, {Gawiser}, {Ivezic}, {Jones}, {Lochner}, {Lund},
  {Mahabal}, {Nidever}, {Olsen}, {Ridgway}, {Rhodes}, {Shemmer}, {Trilling},
  {Vivas}, {Walkowicz}, {Willman}, {Yoachim}, {Anderson}, {Antilogus}, {Angus},
  {Arcavi}, {Awan}, {Biswas}, {Bell}, {Bennett}, {Britt}, {Buzasi},
  {Casetti-Dinescu}, {Chomiuk}, {Claver}, {Cook}, {Davenport}, {Debattista},
  {Digel}, {Doctor}, {Firth}, {Foley}, {Fong}, {Galbany}, {Giampapa}, {Gizis},
  {Graham}, {Grillmair}, {Gris}, {Haiman}, {Hartigan}, {Hawley}, {Hlozek},
  {Jha}, {Johns-Krull}, {Kanbur}, {Kalogera}, {Kashyap}, {Kasliwal}, {Kessler},
  {Kim}, {Kurczynski}, {Lahav}, {Liu}, {Malz}, {Margutti}, {Matheson},
  {McEwen}, {McGehee}, {Meibom}, {Meyers}, {Monet}, {Neilsen}, {Newman},
  {O'Dowd}, {Peiris}, {Penny}, {Peters}, {Poleski}, {Ponder}, {Richards},
  {Rho}, {Rubin}, {Schmidt}, {Schuhmann}, {Shporer}, {Slater}, {Smith},
  {Soares-Santos}, {Stassun}, {Strader}, {Strauss}, {Street}, {Stubbs},
  {Sullivan}, {Szkody}, {Trimble}, {Tyson}, {de Val-Borro}, {Valenti},
  {Wagoner}, {Wood-Vasey}, \& {Zauderer}}]{lsst17}
{LSST Science Collaboration}, {Marshall}, P., {Anguita}, T., {et~al.} 2017,
  arXiv e-prints, arXiv:1708.04058

\bibitem[{{Mallama}(2021)}]{Mallama21}
{Mallama}, A. 2021, arXiv e-prints, arXiv:2101.00374

\bibitem[{{McDowell}(2020)}]{McDowell20}
{McDowell}, J.~C. 2020, \apjl, 892, L36

\bibitem[{{Mr{\'o}z} {et~al.}(2022){Mr{\'o}z}, {Otarola}, {Prince}, {Dekany},
  {Duev}, {Graham}, {Groom}, {Masci}, \& {Medford}}]{Mroz22}
{Mr{\'o}z}, P., {Otarola}, A., {Prince}, T.~A., {et~al.} 2022, \apjl, 924, L30

\bibitem[{{Niino} {et~al.}(2022){Niino}, {Doi}, {Sako}, {Ohsawa}, {Arima},
  {Jiang}, {Tominaga}, {Tanaka}, {Li}, {Niu}, {Tsai}, {Kobayashi}, {Takahashi},
  {Kondo}, {Mori}, {Aoki}, {Arimatsu}, {Kasuga}, \& {Okumura}}]{KISO22}
{Niino}, Y., {Doi}, M., {Sako}, S., {et~al.} 2022, \apj, 931, 109

\bibitem[{{Sako} {et~al.}(2018){Sako}, {Ohsawa}, {Takahashi}, {Kojima}, {Doi},
  {Kobayashi}, {Aoki}, {Arima}, {Arimatsu}, {Ichiki}, {Ikeda}, {Inooka}, {Ita},
  {Kasuga}, {Kokubo}, {Konishi}, {Maehara}, {Matsunaga}, {Mitsuda}, {Miyata},
  {Mori}, {Morii}, {Morokuma}, {Motohara}, {Nakada}, {Okumura}, {Sarugaku},
  {Sato}, {Shigeyama}, {Soyano}, {Tanaka}, {Tarusawa}, {Tominaga}, {Totani},
  {Urakawa}, {Usui}, {Watanabe}, {Yamashita}, \& {Yoshikawa}}]{KISO18}
{Sako}, S., {Ohsawa}, R., {Takahashi}, H., {et~al.} 2018, in Society of
  Photo-Optical Instrumentation Engineers (SPIE) Conference Series, Vol. 10702,
  Ground-based and Airborne Instrumentation for Astronomy VII, ed. C.~J.
  {Evans}, L.~{Simard}, \& H.~{Takami}, 107020J

\bibitem[{{Tregloan-Reed} {et~al.}(2021){Tregloan-Reed}, {Otarola},
  {Unda-Sanzana}, {Haeussler}, {Gaete}, {Colque}, {Gonz{\'a}lez-Fern{\'a}ndez},
  {Anais}, {Molina}, {Gonz{\'a}lez}, {Ortiz}, {Mieske}, {Brillant}, \&
  {Anderson}}]{TregloanReed21}
{Tregloan-Reed}, J., {Otarola}, A., {Unda-Sanzana}, E., {et~al.} 2021, \aap,
  647, A54

\bibitem[{{Tyson} {et~al.}(2020){Tyson}, {Ivezi{\'c}}, {Bradshaw}, {Rawls},
  {Xin}, {Yoachim}, {Parejko}, {Greene}, {Sholl}, {Abbott}, \&
  {Polin}}]{Tyson20}
{Tyson}, J.~A., {Ivezi{\'c}}, {\v{Z}}., {Bradshaw}, A., {et~al.} 2020, \aj,
  160, 226

\bibitem[{{Witze}(2019)}]{Witze19}
{Witze}, A. 2019, \nat, 575, 268

\bibitem[{{Zhang} {et~al.}(2020){Zhang}, {Zhang}, {Qu}, {Wang}, {Zhu}, {Zhu},
  {Zhang}, {Chen}, {Feng}, {Jia}, {Chen}, \& {Wang}}]{Zhang20}
{Zhang}, Y.-h., {Zhang}, H.-f., {Qu}, W.-q., {et~al.} 2020, in Society of
  Photo-Optical Instrumentation Engineers (SPIE) Conference Series, Vol. 11454,
  Society of Photo-Optical Instrumentation Engineers (SPIE) Conference Series,
  1145407



  
\end{thebibliography}

\end{document}